\documentclass[fleqn,twoside]{article}
\usepackage{espcrc2}

\usepackage{graphicx}
\usepackage[figuresright]{rotating}

\newcommand{\AmS}{{\protect\the\textfont2
  A\kern-.1667em\lower.5ex\hbox{M}\kern-.125emS}}

\title{Phototube tests in the MiniBooNE experiment}

\author{Andrew O. Bazarko\footnotemark\address[]{Department of Physics, 
Princeton University, Princeton, New Jersey 08544-0708, USA}
for the BooNE Collaboration and the MiniBooNE PMT group~\cite{pmt_group} 
}
\begin{document}

\begin{abstract}
The MiniBooNE neutrino oscillation experiment at Fermilab uses 
1520 8-inch PMTs: 1197 PMTs are Hamamatsu model R1408 and the 
rest are model R5912.   
All of the PMTs were tested to qualify for inclusion in the detector, sorted 
according to their charge and time resolutions and dark rates.
Seven PMTs underwent additional low light level tests. 
The relative detection efficiency as a function of incident angle for seven 
additional PMTs was measured.  Procedures and results are presented. 
\vspace{1pc}\break
PACS numbers: 85.60.Ha, 14.60.Pq \hfill\break
Keywords: phototube, photomultiplier, MiniBooNE, R1408, R5912
\end{abstract}

\renewcommand\thefootnote{\fnsymbol{footnote}}
\maketitle

\section{Introduction}
\footnotetext{email: bazarko@princeton.edu, \hfill\break
telephone: +1-609-897-8645, \hfill\break 
Presented at Beaune 2005: 4th International Conference on 
New Developments in Photodetection, 
Beaune, France, 19-24 June 2005.}       
The MiniBooNE
experiment at the Fermi National Accelerator Lab 
is searching for neutrino oscillations of the type $\nu_\mu \to \nu_e$ 
\cite{boone}.  
The experiment is designed to confirm or rule out the evidence 
for such oscillations presented
by the LSND experiment, performed at Los Alamos in 1993-1998 
\cite{lsnd01}.
MiniBooNE started running in 2002 and expects to continue taking data at least 
into 2006. 
The MiniBooNE detector is a 12 m diameter spherical tank of undoped
mineral oil.  The main tank volume is defined by an
optical barrier and is viewed by 1280 phototubes.  
Outside the barrier is a veto region viewed by an 
additional 240 tubes.  
Of the 1520 phototubes in the MiniBooNE detector, 1197 are inherited 
from the LSND experiment, and the rest were more recently purchased 
from Hamamatsu.  The LSND tubes are 8 inch (20 cm) in diameter, 
9 stage, Hamamatsu R1408 PMTs, while those newly purchased are
8 inch, 10 stage, Hamamatsu model R5912 PMTs. 

\begin{figure}[tb]
\vspace{9pt}
\includegraphics[width=80mm,bb=0 200 792 600]{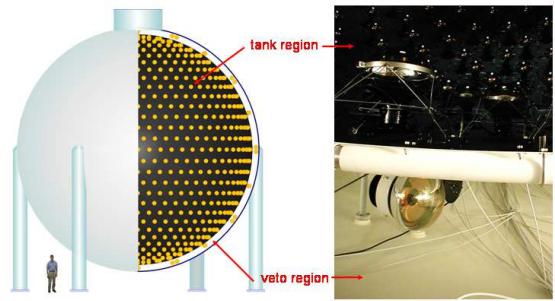}
\caption{Schematic of the MiniBooNE detector with cutaway showing the layout of 
8-inch phototubes in the black main region and in the white veto region.
The photograph of the two regions was taken while the detector was under construction.} 
\label{fig:detector}
\vspace*{-4mm}
\end{figure}

A schematic and photo of the neutrino detector is shown in Fig.~\ref{fig:detector}.  
More details on the tests used to characterize phototubes
for installation in the detector can be found in Ref.~\cite{fleming}, 
and more on the angular tests can be found in 
Ref.~\cite{gladstone}. 

\section{PMT qualification and characterization}
Prior to installation in the MiniBooNE detector, all PMTs were tested to measure:
   dark current,
   time jitter,
   charge resolution,
   double pulsing, and
   pulse shape.
Using these characteristics the PMTs 
were qualified for installation, and 
grouped into four categories according
to time and charge resolution.  Each PMT
category was randomly distributed in the detector.
Those with the worst time resolution but low 
dark rate were placed in the veto region.

\begin{figure}[tb]
\vspace{9pt}
\includegraphics[width=80mm]{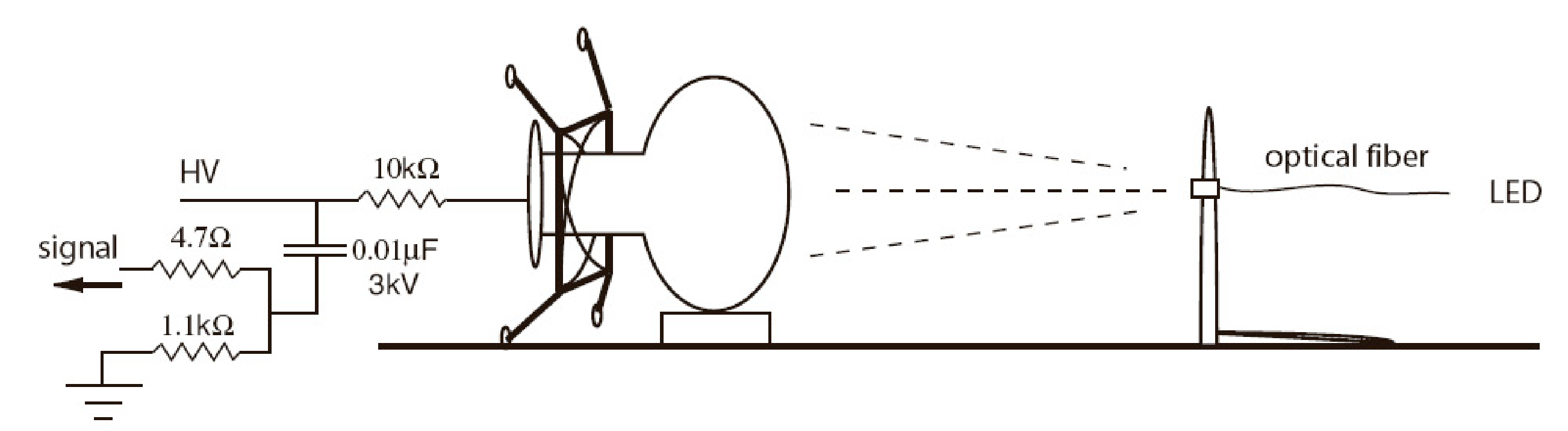}
\vspace*{-13mm}
\caption{Schematic of the PMT test setup.}
\label{fig:led}
\vspace*{-6mm}
\end{figure}

A schematic of a single tube in the test setup is shown in Fig.~\ref{fig:led}.
The bulk of the measurements were performed in a dark room that 
could accommodate up to 46 phototubes at a time.  
Tubes were conditioned under high voltage in the dark for 12-24 hours before testing.
Dark rates were recorded at a range of operating voltages with no light 
source.  Light from an LED pulser was distributed to each tube with optical fibers,
with the end of each fiber positioned about 20 cm from PMT face.
The LED wavelength was 450 nm and it flashed with 1 ns pulses at a rate of 1 kHz, 
producing light levels ranging from 0.5 to 2 photoelectrons (PE)
per pulse.

\begin{figure}[bht]
\vspace{-9pt}
\includegraphics[width=87mm,bb=180 40 575 580,clip]{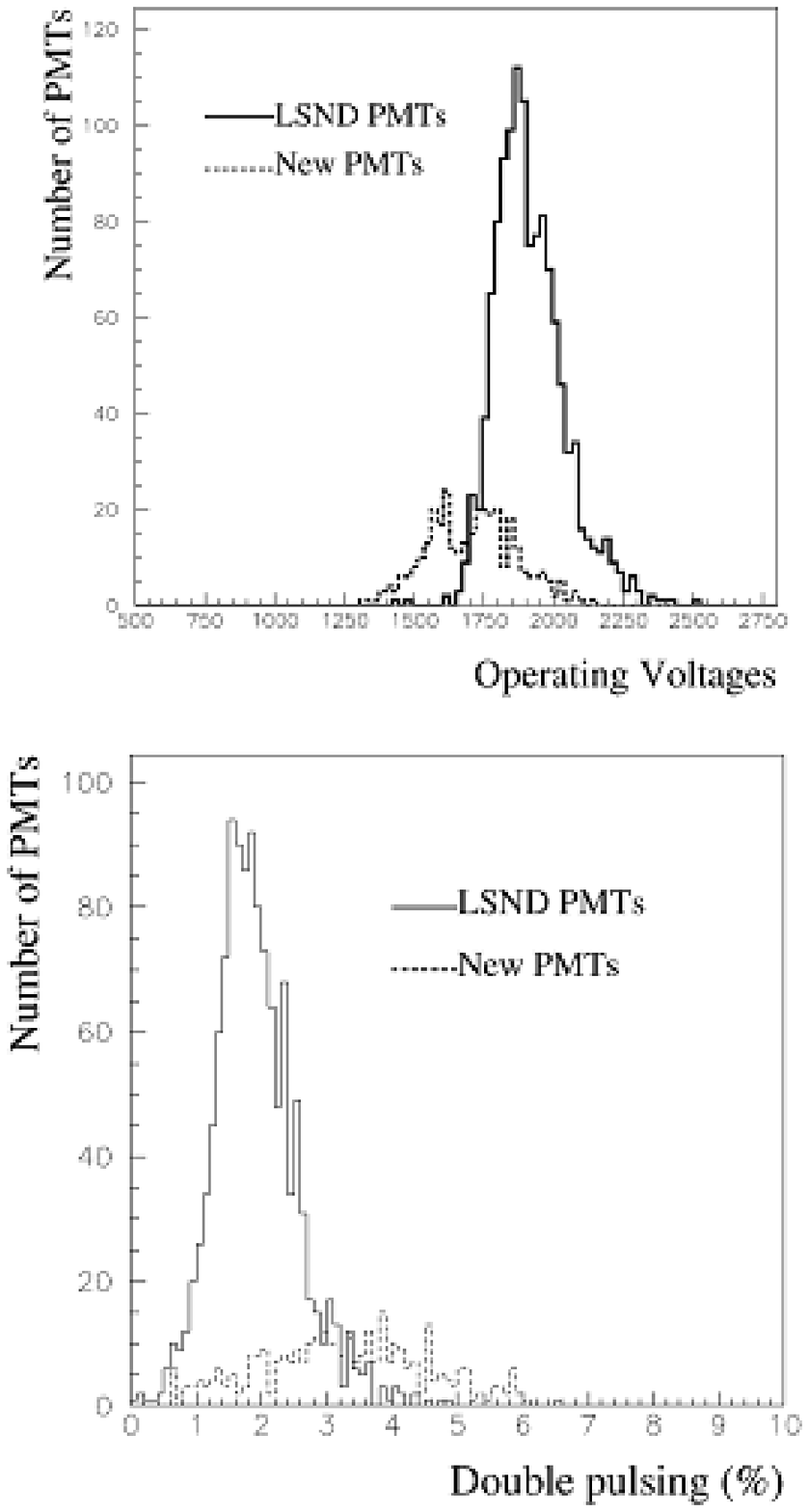}
\vspace{-18pt}
\caption{Operating voltages and post-pulsing rates of R1408 (LSND) and R5912 (``new'')
PMTs prior to selection for installation in MiniBooNE.}
\vspace*{-4mm}
\label{fig:global2}
\vspace*{-4mm}
\end{figure}

Operating voltage was selected to obtain a gain of $1.6\times 10^7$ electrons/PE.
The gain was determined by dividing the average response --- the total charge 
for all pulses with response above threshold divided by the number of 
such responses --- by the average 
number of PEs for the given light level, which was determined from the number of 
null responses.
\begin{figure}[bht]
\vspace{9pt}
\includegraphics[width=87mm,bb=180 40 575 580]{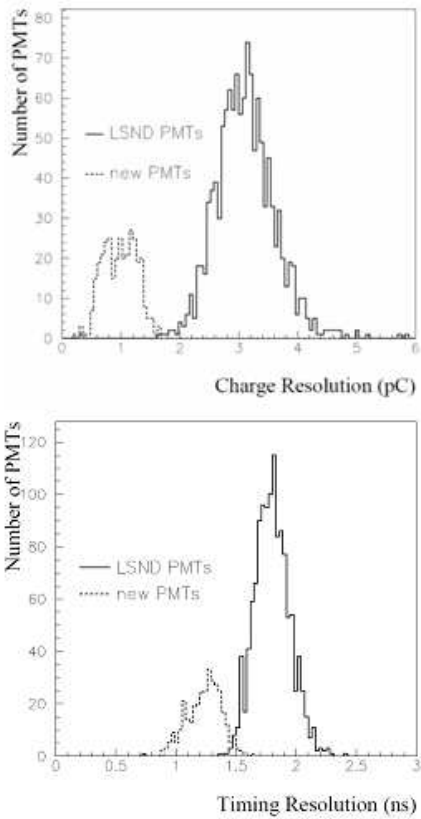}
\vspace*{-10mm}
\caption{Charge and time resolutions of R1408 (LSND) and R5912 (``new'')
PMTs prior to selection for installation in MiniBooNE.}
\label{fig:global1}
\vspace*{-4mm}
\end{figure}
Charge resolution was determined from the width of the one 
photoelectron peak.
Time resolution
was taken as the width of the distribution of the time the PMT pulse
crossed half maximum of its response to the LED pulse.  
In studying 
post-pulsing behavior, the type of most concern for MiniBooNE is
so-called early post-pulsing, when the second
pulse occurs 8-60 ns after the primary pulse --- because data in the 
experiment are recorded in 100 ns intervals.  Such post-pulsing
can occur when an electron from the primary cascade is
ejected from the first dynode and moves inside the PMT dome before settling 
back on the first dynode to initiate a secondary cascade.  Tubes were
found to have such double pulse rates of a few percent.
Distributions of these measurements are shown in Figs.~\ref{fig:global2}
and~\ref{fig:global1}. 

\section{Low light level tests}

Seven phototubes (three R5912 and four R1408) were measured 
using very low light levels.  The same apparatus as described
above was used with the addition of neutral density filters so that 
the probability for 
producing 2 PE or more was less than 0.001, or equivalently less than
45 visible PMT response waveforms per 1000 LED triggers. 

\begin{figure}[bht]
\vspace{-9pt}
\includegraphics[width=80mm,bb=50 180 750 580]{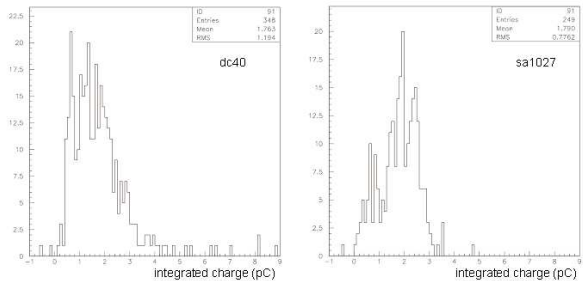}
\caption{Integrated charge distributions for single photoelectron response
 for one R1408 (dc40) and one R5912 (sa1027) PMTs .}
\label{fig:low_light}
\vspace*{-4mm}
\end{figure}

The two types of PMTs show differences in single photoelectron response.  
The integrated charge distributions from 
one R1408 PMT and one R5912 PMT are shown in Figure~\ref{fig:low_light}. 
The three measured R5912 tubes had average charge 
response of 2.26 pC and average charge distribution width of 0.89 pC, 
while the four R1408 tubes had average charge of 1.96 pC and average
charge width of 1.32 pC. The R1408 tubes have a pronounced high charge 
tail.

\section{Incident angle dependence}

The apparatus used to measure the dependence of the PMT
response on the angle of the incident light
consisted of 
a 150 liter tank with walls 
coated with the same black paint used in MiniBooNE.
A 26 cm diameter window allowed light from outside
to illuminate a PMT housed inside the tank, and the PMT could be
rotated in pitch and yaw with externally-coupled mechanical 
controls.
An LED light source was directed toward the window from 
a distance of 3 m so that the entire face of the tube was
approximately uniformly illuminated
with parallel light.  
Ambient magnetic fields were suppressed by a $\mu$-metal shield outside
the tank.  
Measurements were performed with and without mineral oil in the tank. 

\begin{figure}[tbh]
\vspace{9pt}
\includegraphics[width=70mm,bb=150 45 574 582]{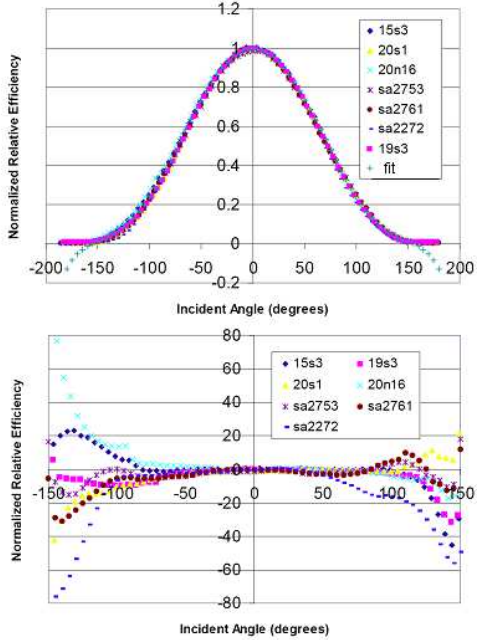}
\caption{Top: Response versus incident angle for seven PMTs in mineral oil.  Each tube's
  response is shown relative to its maximum response, which is assumed
  to occur at incident angle zero.  The symmetric polynomial fit to these
  data is
  also shown.  Bottom:  The percentage deviation of each tube's response from
  the fit. }
\label{fig:angle_1}
\vspace*{-4mm}
\end{figure}

\begin{figure}[tbh]
\vspace{36pt}
\includegraphics[width=70mm,bb=45 60 1250 852]{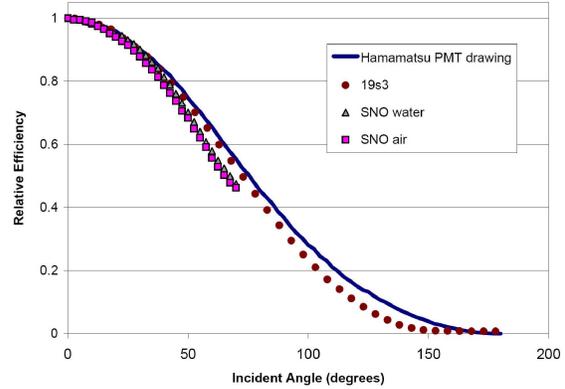}
\caption{Response versus incident angle relative to the maximum at zero incidence
  angle.  The prediction based solely on the PMT geometrical shape is compared
with the MiniBooNE measurement in oil (tube 19s3 shown) and measurements in air and water from SNO.}
\label{fig:angle_4}
\vspace*{-4mm}
\end{figure}

Response measurements were performed by varying the pitch angle from 
$-180^\circ$ to $+180^\circ$ in $5^\circ$ steps, and four R1408 and three
R5912 tubes were tested.  Each phototube is rotationally symmetric about its 
central vertical axis, except for its dynode structure.  The R1408 tubes have
``venetian blind'' dynodes and the R5912 tubes have ``box and line'' dynodes. 
No significant differences were observed when the dynodes 
were oriented horizontally, vertically, or at $45^\circ$ relative to the
pitch angle scan.
In addition, one tube's response as a function of yaw angle  
was compared to its response in pitch angle
and no significant difference was 
observed.  

Maximum response was assumed to occur at zero incident angle, and
the relative response measured in oil for seven phototubes is shown in
Fig.~\ref{fig:angle_1}.
A symmetric polynomial fits these data well.

The relative response is predominantly due simply to the
solid angle subtended by the tube as a function of incident angle. 
The curve labeled ``Hamamatsu PMT drawing'' in Fig.~\ref{fig:angle_4} 
indicates the relative response predicted using the PMT shape  
specified in the Hamamatsu technical drawing, which is 
nearly hemispherical but a bit more bulbous.
The medium surrounding the tube also plays a role:  
Fig.~\ref{fig:angle_4} shows that the relative response progressively 
increases for air, water, and mineral oil, which is due to the progressively
better matching of the medium's refractive index with that of the PMT
glass.  (The air and water measurements shown are from the SNO 
experiment\cite{sno}, which employs R1408 tubes.  MiniBooNE's air measurements
agree with those from SNO.)  Finally, Fig.~\ref{fig:angle_4} indicates
the difference that remains between the measured relative response in oil and that
predicted due to the PMT geometry.  This information is used to model 
the PMT response in the MiniBooNE detector simulation.




\end{document}